\documentclass[12pt]{article}
\usepackage{amsmath,amssymb,amscd,latexsym,euscript,mathrsfs}
\usepackage[all]{xy}

\newcommand{\coLim}{\underrightarrow{\lim}}

\newcommand{\Set}{{\rm Set}}
\newcommand{\Ab}{{\rm Ab}}
\newcommand{\Ob}{{\rm Ob\,}}
\newcommand{\Mor}{{\rm Mor\,}}

\newcommand{\pt}{{\,\rm pt}}

\newcommand{\mC}{{\mathscr C}}
\newcommand{\mN}{{\,\mathcal N}}

\newcommand{\ZZ}{{\,\mathbb Z}}
\newcommand{\II}{{\,\mathbb I}}
\newcommand{\fP}{{\mathfrak{P}}}

\newtheorem{theorem}{\bf Theorem}[section]
\newtheorem{lemma}[theorem]{\bf Lemma}
\newtheorem{proposition}[theorem]{\bf Proposition}

\newtheorem{definition}{\sc Definition}[section]
\newtheorem{example}[definition]{\sc Example}

\def\leq{\leqslant}
\def\geq{\geqslant}

\begin{document}

%\begin{flushleft}
%{УДК 512.66:519.7
%}
%\end{flushleft}

\begin{center}
 {\large HOMOLOGY GROUPS OF PIPELINE PETRI NETS
%Homology groups of pipeline Petri nets
}\footnote{\rm This work was performed as a part of the Strategic Development 
Program at the National Educational Institutions of the Higher Education, N 2011-PR-054}
\end{center}

\begin{center}
A.A. Husainov, E.S. Bushmeleva, T.A. Trishina
\end{center}

\begin{abstract}
We study homology groups of elementary Petri nets for the pipeline systems.
We show that the integral homology groups of these nets  equal the
group of integers in dimensions $0$ and $1$, and they are zero in other dimensions. We prove that directed
homology groups of elementary Petri nets are zero in all dimensions.
\end{abstract}

2000 Mathematics Subject Classification 18G10, 18G35, 55U10, 68Q10, 68Q85


\section*{Introduction}

The homology groups of elementary Petri nets were introduced in \cite{X20042}.
In \cite{X20042}, it was built also an algorithm to compute the first homology group of
asynchronous systems, based on which we calculated the homology of Petri net of pipeline $\fP_3$,
consisting of three transitions. In \cite{X2012}, an algorithm for computing the homology
groups of elementary Petri net was developed.
An example of calculating the homology groups of elementary Petri net for pipeline $\fP_4$
$$
 
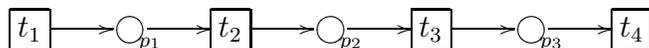
\xymatrix{
*+[F]{t_1}\ar[r] & *+<9pt>[o][F]{~} \ar[r]_(0.2){p_1} & *+[F]{t_2} \ar[r] & *+<9pt>[o][F]{~} \ar[r]_(0.2){p_2} & *+[F]{t_3}
\ar[r] & *+<9pt>[o][F]{~}
\ar[r]_(0.2){p_3} & *+[F]{t_4}
}
$$
was considered.

We calculated the homology group for
 $\fP_n$, with $n=2, 3, 4, 5$ by the software described in \cite{bus2012}. 
In these cases, the homology group for the pipelines are $H_0(\fP_n)=H_1(\fP_n)=\ZZ$,
and $H_k(\fP_n)=0$ for $k\geq 2$.
There was a conjecture that this is true for all $n\geq 2$.
 In this paper, we prove this conjecture. In addition, the second co-author of this
 paper has developed the software designed to calculate the directed homology groups
 of the state spaces. Calculations showed that the directed homology groups 
of elementary Petri nets
 $\fP_n$ for $n=2,3,4,5$ are equal to zero in all dimensions. In this paper, we
 show that this is true for all $n\geq 2$.

\section{Preliminaries}

For any category $\mC$ denote by $\Ob\mC$ the class of its objects,
$\Mor\mC$ the class of all morphisms, and $\mC^{op}$ the opposite category.
For any $A,B\in \Ob\mC$ the set of all morphisms $A\to B$ in a category $\mC$ is denoted by $\mC(A, B)$.

\subsection{Partial maps} 
For any sets $S$ and $T$ {\em a partial map} $f: S\rightharpoonup T$
is a relation $f\subseteq S\times T$ with the following property:
$$
 (\forall s\in S)(\forall t_1, t_2\in T) ~(s, t_1)\in f ~\& ~ (s,t_2)\in f ~ \Rightarrow~ t_1=t_2\,.
$$
We denote $PSet$ category of sets and partial maps.

For any set $S$ denote $S_*=S\sqcup\{*\}$.
Let $\Set_*$ be a category of sets of the form  $S_*$ 
where $S$ are arbitrary sets.
For any of its objects $S_*$ and $T_*$, let the set of morphisms $\Set_*(S_*, T_*)$
consists of maps $f: S_*\to T_*$ satisfying the equality $f(*)=*$.
It is easy to see that functor $\Phi: PSet\to \Set_*$, defined on objects as
$\Phi(S)=S_*$, and on morphisms
$$
\Phi(f)(x)= \left\{
\begin{array}{cl}
f(x), & \mbox{ if } f(x) \mbox{ defined},\\
*, & \mbox{ if } f(x) \mbox{ is not defined or } x=*,\\
\end{array}
\right.
$$
will provide an isomorphism of categories $PSet$ and $\Set_*$.
This isomorphism allows us to identify
category $PSet$ with  $\Set_*$
and we can consider a partial map as the (total) function preserving point $*$.

\subsection{Partial action of a monoid} 
Suppose $M$ be a monoid. The operation in $M$ is denoted by $\cdot_M$.
Each monoid $M$ will be considered as a small category that has
single object $\pt_M$.  The set of morphisms of this category is the set 
$M$.
In particular, the opposite monoid $M^{op}$ is defined. Product of the elements $\mu_1, \mu_2\in M$ in this monoid is given by $\mu_1\cdot_{M^{op}}\mu_2= \mu_2\cdot_{M}\mu_1$.
 {\em Partial right action of a monoid $M$ on the set $S$}
is an arbitrary homomorphism $M^{op} \to PSet(S,S)$.
In accordance with the above agreement,
a partial action will be viewed as a homomorphism $M^{op}\to \Set_*(S_*,S_*)$.
It follows that it can be regarded as
 a functor $M^{op}\to \Set_*$.

\subsection{Trace monoids and state space} 
Let  $E$ be a set, and let $I\subseteq E\times E$ be a irreflective symmetric relation on $E$.
Let $M(E,I)$ be a monoid generated by the set $E$ and given 
by the equalities $a b= b a$ for all $(a,b)\in I$. The monoid $M(E,I)$
is called to be a {\em  free partially commutative monoid} or {\em trace monoid}.
$M(E,I)$ is equal to the quotient monoid $E^*/\equiv$ where $E^*$ is the monoid of words
and $\equiv$ is the smallest 
congruence relation containing a pair of $(ab,ba)$ for all pairs $(a,b)\in I$.
{\em The space of states} is a pair $(S,M(E,I))$,  where $M(E,I)$ is a trace
monoid with action on a set $S$.

\subsection{The state space of elementary Petri nets}
For any elementary Petri net $\mN=(P, E, pre, post, s_0)$, we consider a 
trace monoid $M(E,I)$, generated by the set $E$ with the
independence relation
$$
(a,b)\in I ~\Leftrightarrow~ (pre(a)\cup post(a)) \cap (pre(b)\cup post(b))= \emptyset.
$$

Suppose $a\in E$. For any subset $s\subseteq P$,
with properties
\begin{itemize}
\item $pre(a)\subseteq s$,
\item $(s\setminus pre(a))\cap post(a)=\emptyset$,
\end{itemize}
we denote $s\cdot a= (s\setminus pre(a))\cup post(a)$.
This defines for each $a\in E$ a partial map $s\mapsto s\cdot a$, 
of the set $\{0,1\}^P$ in  $\{0,1\}^P$. 
Here the subsets $s\subseteq P$ are considered as their characteristic functions.

The obtained partial map (corresponding to the relation of the transition from \cite{win1995}), 
can be defined as the set of triples $(s, a, s')$ with the property
$$
pre(a)\subseteq s ~ \&~ post(a) \subseteq s' ~
\&~ s\setminus pre(a)= s'\setminus post(a).
$$
Using the induction on the length of the trace $\mu= a_1 a_2\cdots a_n$ by the formula
$s\cdot a_1 a_2 \cdots a_n= (s\cdot a_1 a_2\cdots a_{n-1})\cdot a_n$,
we can define a partial right action of a monoid $M(E,I)$ on $\{0,1\}^P$.
The resulting state space $(\{0,1\}^P, M(E,I))$ will be called a
{\em state space of elementary Petri nets}.

If we consider only the set of reachable states instead of $\{0,1\}^P$, then this partial
action will define the {\em space of reachable states of the elementary Petri net}.

\section{Homology of semicubical sets and state spaces}

Recall the definition of semicubical set and its homology groups \cite{X20081}.
We describe semicubical set corresponding to a state category with the
same homology groups.

\subsection{Semicubical sets}

A {\em semicubical set} $X=(X_n, \partial_i^{n,\varepsilon})$ is given as a
sequence
 sets $X_n$, $n\geq 0$ with a family of maps
$\partial_i^{n,\varepsilon}: X_n\to X_{n-1}$, defined by $1\leq i\leq n$,
$\varepsilon\in \{0,1\}$, for which the following diagrams
$$
\xymatrix{
		X_n \ar[d]_{\partial_j^{n,\beta}} \ar[rr]^{\partial_i^{n,\alpha}}
&& X_{n-1} \ar[d]^{\partial_{j-1}^{n-1,\beta}}\\
		X_{n-1} \ar[rr]_{\partial_i^{n-1,\alpha}} && X_{n-2}
}
$$
are commutative
for all $n\geq 2, ~ 1\leq i<j \leq n$.

{\em Morphism of semicubical sets $f: X\to \tilde{X}$}
is a sequence of maps
$f_n: X_n \to \tilde{X}_n$, $n\geq 0$, for which the diagrams
$$
\xymatrix{
	X_n \ar[rr]^{f_n} \ar[d]_{\partial_i^{n,\varepsilon}} && \tilde{X}_n \ar[d]^{\tilde{\partial}_i^{n,\varepsilon}}\\
	X_{n-1} \ar[rr]_{f_{n-1}} && \tilde{X}_{n-1}
}
$$
are commutative for all $1\leq i\leq n$ and $\varepsilon\in \{0,1\}$.
If the morphism semicubical sets consist of inclusions $X_n\subseteq \tilde{X}_n$
for all $n\geq 0$, then $X$ is called to be a {\em semicubical subset} of $\tilde{X}$.

\subsection{Homology of semicubical sets}
Let $X= (X_n, \partial_i^{n,\varepsilon})$ be a semicubical set.
Consider the complex
$$
	0\leftarrow LX_0\stackrel{d_1}\leftarrow LX_1\stackrel{d_2}\leftarrow LX_2\stackrel{d_1}\leftarrow \cdots,
$$
consisting of free Abelian groups $LX_n$, $n\geq 0$, generated by the sets $X_n$, and differentials,
acting on the basis elements by the formula
$$
d_n (\sigma) = \sum\limits_{i=1}^n (-1)^i (\partial_i^{n,1}(\sigma)-\partial_i^{n,0}(\sigma)).
$$
The homology groups of this complex are
called to be {\em homology groups $H_n(X)$ of semicubical set $X$}, $n\geq 0$.

\begin{proposition}\label{exactcube}
Suppose that $X=X_1\cup X_2$ is a union of semicubical subsets.
Then there is a long exact sequence 
of groups
\begin{multline*} %\label{exactcube}
  0 \leftarrow H_0(X) \leftarrow H_0(X_1)\oplus H_0(X_2)
\leftarrow H_0(X_1\cap X_2) \leftarrow \cdots \\
  \cdots \leftarrow H_n(X) \leftarrow H_n(X_1)\oplus H_n(X_2)
\leftarrow H_n(X_1\cap X_2) \leftarrow \cdots \\
\end{multline*}
\end{proposition}
{\sc Proof.}
Consider the complex $C_n(X)=LX_n$ with differentials
$d_n (\sigma) = \sum\limits_{i=1}^n (-1)^i (\partial_i^{n,1}(\sigma)-\partial_i^{n,0}(\sigma))$. 
Consider the exact sequence of 
complexes associated with the homomorphism $\sigma_1\oplus\sigma_2\mapsto \sigma_1-\sigma_2$,
$$
	0 \rightarrow C_n(X_1\cap X_2) \stackrel{\theta_n}\rightarrow
C_n(X_1)\oplus C_n(X_2)
  \stackrel{-}\rightarrow C_n(X_1\cup X_2) \to 0,
$$
where $\theta_n(\sigma)= \sigma\oplus\sigma$ for each $\sigma\in (X_1\cap X_2)_n$.
A long exact sequence corresponding to this short exact sequence,
will be required.
\hfill $\Box$

\subsection{Homology of state category and semicubical sets}
Let $(S, M(E,I))$ be a state space.
Consider an arbitrary linear order relation on $E$.
It defines a semicubical set
\begin{multline*}
	Q_n(S, E, I) = \{(s, a_1, \ldots, a_n)\in S\times E^n ~|
			~ a_1<\cdots <a_n ~\&~ s\cdot a_1\cdots a_n\in S \\
			~\&~ (a_i, a_j)\in I \mbox{ for all } 1\leq i< j \leq n\}, ~~ n\geq 0,
\end{multline*}
with boundary operators
$$
\partial_i^{n, \varepsilon}(s, a_1, \ldots, a_n)= (s\cdot a_i^{\varepsilon}, a_1, \ldots, a_{i-1}, a_{i+1}, \ldots, a_n),
$$
for $1\leq i\leq n$, $\varepsilon\in \{0,1\}$. There $a_i^0=1$ and $a_i^1=a_i$.

Let $(S, M(E,I))$ be a state space. Its {\em state category $K(S)$} is defined as a small category with objects $s\in S$.
Morphisms $s\stackrel{\mu}\to t$ are defined as triples of elements $s,t\in S$ and $\mu\in M(E,I)$ 
satisfying  $s\cdot\mu=t$.
Composition is given by the formula $(t\stackrel{\nu}\to u)\circ (s\stackrel{\mu}\to t)=(s\stackrel{\mu\nu}\to u)$.

For any elementary Petri net $\mN$, we consider the state space and its state category. 
Let $K(S)$ be a full subcategory of the state category consisting of reachable states. 
 {\em Homology groups $H_n(\mN)$} is defined as homology groups $H_n(K(S))$ 
of the nerve of the category $K(S)$.
According to \cite[Corollary 4]{X2012}, $H_n(K(S))\cong H_n(Q(S,E,I))$, for all $n\geq 0$.
Therefore, the homology groups of elementary Petri nets $H_n(\mN)$ are isomorphic to the homology of semicubical
sets corresponding to its space of reachable states.

\section{State category of the pipeline Petri net and its homology groups}

In this section, we will explore the state category  of pipeline elementary Petri net 
and calculate the integral homology groups of this category.

\subsection{State category of pipeline elementary Petri net}

We consider pipeline Petri net $\fP_n$:
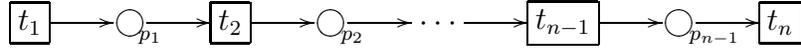
$$
 \xymatrix{
*+[F]{t_1}\ar[r] & *+<9pt>[o][F]{~} \ar[r]_(0.2){p_1} & *+[F]{t_2} \ar[r] & *+<9pt>[o][F]{~} \ar[r]_(0.2){p_2}
& \cdots \ar[r] & *+[F]{t_{n-1}}
\ar[r] & *+<9pt>[o][F]{~}
\ar[r]_(0.35){p_{n-1}} & *+[F]{t_n} \, .
}
$$
Represent the state category of this net as union of two 
 partially ordered sets, each of which has the least 
and the greatest element.

Let $\mN_n$ be the following Petri net
$$
 
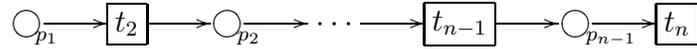
\xymatrix{
*+<9pt>[o][F]{~} \ar[r]_(0.2){p_1} & *+[F]{t_2} \ar[r] & *+<9pt>[o][F]{~} \ar[r]_(0.2){p_2}
& \cdots \ar[r] & *+[F]{t_{n-1}}
\ar[r] & *+<9pt>[o][F]{~}
\ar[r]_(0.35){p_{n-1}} & *+[F]{t_n} \, .
}
$$
Denote by $\mC_n$ its the state category. Every partially ordered set can be considered
as a small category $\mC$ such that for any objects $x, y\in \Ob\mC$, the set $\mC(x,y)$
contains no more than one element, and if the conjunction of $\mC(x,y)\not=\emptyset$ and
$\mC(y,x)\not=\emptyset$ implies $x=y$.

Any state of elementary Petri net $\mN_n$ is given by an arbitrary function
$f: \{p_1, p_2, \cdots, p_{n-1}\}\to \{0, 1\}$. It follows that the state
  can be given by a sequence of
values  $\varepsilon_1 \varepsilon_2 \cdots \varepsilon_{n-1}$, where $\varepsilon_i=f(p_i)$ of $f$
for $1\leq i\leq n-1$. Let $S_1$ the set of states beginning with $\varepsilon_1=1$,
and $S_0$ the set of states, beginning with $\varepsilon_1=0$. The set of all states will be
equal to $S= S_1 \sqcup S_0$.
In general, the elements of $S_1$ and $S_0$ are connected by transitions
$10\varepsilon_3\cdots \varepsilon_{n-1}\stackrel{t_2}\to
01\varepsilon_3\cdots \varepsilon_{n-1}$.

For example, if $n=4$, then $S_1$ and $S_0$ are linearly ordered sets 
corresponding columns of the diagram:

$$
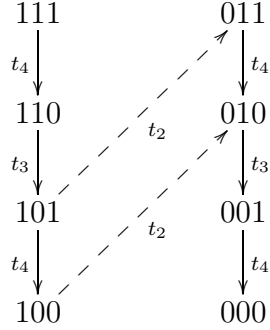
\xymatrix{
111\ar[d]_{t_4} & & 011\ar[d]^{t_4}\\
110\ar[d]_{t_3} & & 010\ar[d]^{t_3}\\
101\ar[d]_{t_4} \ar@{-->}[rruu]_{t_2} & & 001\ar[d]^{t_4}\\
100 \ar@{-->}[rruu]_{t_2}& & 000
}
$$
Transitions between $S_1$ and $S_0$ is shown by the dashed arrows.

For any state $x=\varepsilon_1\cdots\varepsilon_{n-1}$, we introduce a number
$|x|=\varepsilon_1\cdot 2^{n-2}+ \varepsilon_2\cdot 2^{n-3}+ \cdots
+\varepsilon_{n-1}\cdot 2^{0}$. For example, $|11\cdots 1|= 2^{n-1}-1$ and $|00\cdots 0|=0$.
If there is a transition $x \stackrel{t_k}\to y$,
then $|x|<|y|$

\begin{lemma}\label{subnet1}
Category $\mC_n$ is a partially ordered set having the greatest and least elements.
\end{lemma}
{\sc Proof.}
By induction, we prove that $\mC_n$ is a partially ordered set with the least
element of $11\cdots 1$ and the greatest element $00\cdots 0$.
Let it proved for $n-1$. Then  $\mC_{n-1}$ is the partially ordered set with the least
and greatest elements. It is easy to see that the full subcategory of  $\mC_n$ with set of
objects $S_1$ and $S_0$ are isomorphic to the category $\mC_{n-1}$. So, they will be partially ordered
sets. We denote these posets by $S_1$ and $S_0$.
In $S_1$, a least element of the induction assumption is $11\cdots 1$, and the greatest element equals  $10\cdots 0$.
In $S_0$, a least element is $01\cdots 1$, and the greatest equals $00\cdots 0$.

Let $x\mu= y$, $x\in S_1$, $y\in S_0$. Then $\mu$ contains the letter $t_2$.
Therefore, there exist $\mu_1$ and $\mu_2$, such that $\mu = \mu_1 t_2\mu_2$, and $\mu_1$
does not contain $t_2$.
Now, we will rearrange the letters contained in $mu_1$ with the letter $t_2$, until we do not meet $t_3$.
%
% We shall be moved from $t_2$ letters of $\mu_1$, as long as
%not meet $t_3$.
For some $\mu_1$ and $\mu_2$, we obtain the decomposition
$\mu= \mu_1 t_3 t_2 \mu_2$.

The following step we will perform until we get the decomposition of the form
$\mu= t_k\cdots t_2 \mu'$.

Suppose that $\mu= \mu_1 t t_{k-1} \cdots t_2$, for some  $k\geq 3$ and $t\in E$.
We describe the action taken by us in each of the following possible cases:

\begin{enumerate}
\item $t=t_k$ $\Rightarrow$ increase $k$ by $1$,
\item $t=t_i$, for some $i>k$ $\Rightarrow$ swap $t$ with $t_{k-1}$,
$t_{k-2}$, $\cdots$, $t_2$.
\end{enumerate}

For $i\leq k-1$, a decomposition of the form $\mu= \mu_1 t_i t_{k-1}\cdots t_2$
 impossible, since an element $x' t_{k-1}\cdots t_2$ is defined if and only if
 $x'=11\cdots 10\varepsilon_{k}\cdots\varepsilon_{n-1}$.
Such element $x'$ can not be obtained after the action of $t_i$
for $2\leq i\leq k-1$.
It follows that the iteration of the described action will lead to the decomposition
$\mu= t_k\cdots t_2 \mu'$.

Now we will prove that $x\mu=x\nu=y\in S$ implies $\mu=\nu$,
which implies that $\mC_n$ is a preordered set.
%========================================================

If $x\mu=x\nu=y\in S_0$ and $y\in S_0$, then $\mu=\nu$ because of $S_0$ is a partially
ordered set. The same is true if $x\mu=x\nu=y\in S_1$ and $y\in S_1$.
There are no $\mu\in M(E,I)$ and $x\in S_0$, for which $x\mu\in S_1$.

Now consider the case $x\mu=x\nu=y\in S_0$, $x\in S_1$.
In this case, there exist decompositions $\mu= t_k\cdots t_2 \mu'$ and
$\nu= t_m\cdots t_2 \nu'$.
Since $x t_k t_{k-1}\cdots t_2 \mu' \in S_0$, then
$x=11\cdots 10\varepsilon_{k+1}\cdots \varepsilon_{n-1}$, for some
$\varepsilon_{k+1}$, $\cdots,$ $\varepsilon_{n-1}\in \{0,1\}$.
Hence $m=k$. We get the equality
$$
x t_k\cdots t_2 \mu' = x t_k\cdots t_2 \nu'
$$
Since $x t_k\cdots t_2 \in S_0$ and $S_0$ -- partially ordered set
it follows from this equation $\mu'=\nu'$.
Therefore, $\mu=\nu$ and $\mC_n$ -- preordered set.
If for some $x=\varepsilon_1\cdots\varepsilon_{n-1}$,
$y= \delta_1\cdots \delta_{n-1}$, there exists $\mu\not=1$, which
$x\mu =y$, then $|x|>|y|$. Therefore, from $x\mu=y$ and $y\nu=x$
will follow the $\mu=\nu=1$. Hence otnshenie preorder will antisymmetric
and $\mC_n$ is partially ordered set.
\hfill $\Box$

\subsection{The homology groups of pipeline}

Let $\fP_n$ -- elementary net pipeline. Delete event $t_1$, we have a network
$\mN_n$. Remove from $\fP_n$ event $t_2$, we get the following elementary network:

$$
 
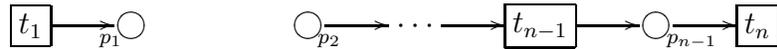
\xymatrix{
*+[F]{t_1}\ar[r]_(0.8){p_1} & *+<9pt>[o][F]{~} &  & *+<9pt>[o][F]{~} \ar[r]_(0.2){p_2}
& \cdots \ar[r] & *+[F]{t_{n-1}}
\ar[r] & *+<9pt>[o][F]{~}
\ar[r]_(0.35){p_{n-1}} & *+[F]{t_n} \, .
}
$$

We denote it by $\mN'_n$.

\begin{proposition}\label{union}
Semicubical set $Q(\fP_n)$ is the union of subsets semicubical
$Q(\mN_n)\cup Q(\mN'_n)$.
Intersection of $Q(\mN_n)\cap Q(\mN'_n)$ semicubical be set with a
two components, each of which is isomorphic $Q(\mN_{n-1})$.
\end{proposition}
{\sc Proof.} There are $Q_0(\fP_n)= Q_0(\mN_n)\cup Q_0(\mN'_n)$,
because all of these sets are $S= \{0,1\}^{n-1}$. If $k\geq 1$,
then since $t_1$ and $t_2$ dependent, they can not belong
a set of mutually independent events $(e_1, \cdots, e_m)$.
This means that for every $(s, e_1, \cdots, e_m)\in Q_m(\fP_n)$ will be a
$t_1\notin \{e_1, \cdots, e_m\}$ or $t_2\notin \{e_1, \cdots, e_m\}$,
where $(s, e_1, \cdots, e_m)\in Q_m(\mN_n)\cup Q_m(\mN'_n)$.
Because of $(s, e_1, \cdots, e_m)\in Q_m(\mN_n)$ follows
$(s\cdot e_i^{\varepsilon}, e_1, \cdots, e_{i-1}, e_{i+1}, \cdots, e_m)\in Q_{m-1}(\mN_n)$
and from $(s, e_1, \cdots, e_m)\in Q_m(\mN'_n)$ follows
$(s\cdot e_i^{\varepsilon}, e_1, \cdots, e_{i-1}, e_{i+1}, \cdots, e_m)\in Q_{m-1}(\mN'_n)$,
attachments commute with the boundary operators. So, $Q(\fP_n)$ equals the union
its cubic subsets of $Q(\mN_n)$ and $Q(\mN'_n)$.
\hfill $\Box$

\begin{theorem}
$H_0(\fP_n)=H_1(\fP_n)=\ZZ$, and $H_k(\fP_n)=0$ at $k>1$.
\end{theorem}
{\sc Proof.} By Proposition \ref{union}, $Q(\fP_n)= Q(\mN_n)\cup Q(\mN'_n)$.
Apply Proposition \ref{exactcube}.
By Lemma \ref{subnet1} category network states $\mN_n$ has
initial and terminal object.
Let $\II=\{0,1\}$ -- partially ordered set consisting of the elements of $0$
and $1$, with the usual order relation $0<1$.
It is easy to see that the category of network conditions
$\mN'_{n}$ is isomorphic to the product of $\II\times \mC_{n-2}$. Therefore it has
initial and terminal facilities too. From this follow isomorphisms
  $H_k(Q(\mN_n))=H_k(Q(\mN'_n))=0$, for $k>0$.
The exact sequence \ref{exactcube} leads to the isomorphisms
$H_k(Q(\fP_n))\cong H_{k-1}(Q(\mN_n)\cap Q(\mN'_n))$, for $k\geq 2$.
Of Proposition \ref{union} follows $Q(\mN_n)\cap Q(\mN'_n)\cong Q(\mN_{n-1})\sqcup Q(\mN_{n-1})$.
We get $H_k(Q(\fP_n))\cong H_{k-1}(Q(\mN_{n-1}))\oplus H_{k-1}(Q(\mN_{n-1})) =0$, where
$k\geq 2$.
Consequently, $H_k(\fP_n)=0$ for $k\geq 2$.
Of Proposition \ref{exactcube} also obtain accurate
sequence
\begin{multline*}
	0 \leftarrow H_0(Q(\fP_n)) \leftarrow H_0(Q(\mN_n))\oplus H_0(Q(\mN'_n))\\
		\leftarrow H_0 (Q(\mN_n)\cap Q(\mN'_n)) \leftarrow
H_1(Q(\fP_n)) \leftarrow 0.
\end{multline*}
Group $H_0$ freely generated connected components semicubical sets.
Homomorphism
$$
H_0(\theta): H_0(Q(\mN_n)\cap Q(\mN'_n)) \to H_0(Q(\mN_n)) \oplus H_0(Q(\mN'_n))
$$
induced chain homomorphism
$$
\theta_k: C_k(Q(\mN_n)\cap Q(\mN'_n)) \to C_k(Q(\mN_n)) \oplus C_k(Q(\mN'_n)),
$$
defined by $\sigma \in (Q(\mN_n)\cap Q(\mN'_n))_k$ by the formula
$\theta_k(\sigma)= \sigma\oplus \sigma$.
This implies that $H_0(\theta)$ acts on the homology classes of the formula
$H_0(\theta)(cls(\sigma))= cls(\sigma)\oplus cls(\sigma)$. Because these classes homology
are connected components semicubical sets, then $H_0(\theta)$
 is a homomorphism $\ZZ\oplus\ZZ \to \ZZ\oplus\ZZ$, given by the matrix
$
\left(
\begin{array}{cc}
1 & 1\\
1 & 1
\end{array}
\right)$.
 But $H_1(\fP_n)$ is isomorphic to kernel of the homomorphism $H_0(\theta)$,
and $H_0(\fP_n)$ -- cokernel.
Hence, for example, with a cast of this matrix to Smith normal form, we obtain
$H_0(\fP_n)= H_1(\fP_n)= \ZZ$.
\hfill $\Box$

\section{Directed homology}

We present supporting information on the homology groups Goubault semicubical sets.
We study the properties of directed homology category
states and their interpretation in dimension $0$. Calculate the direction of
homology Petri net pipeline.

\subsection{Homology groups of Goubault}

Let $X= (X_n, \partial^{n,\varepsilon}_i)$ semicubical set.
For all $\varepsilon\in \{0, 1\}$, consider the chain complex of abelian groups
$C_n(X)=L(X_n)$ with the differentials
$$
0 \leftarrow C_0(X) \stackrel{d_1^{\varepsilon}}\leftarrow C_1(X) \leftarrow \cdots
C_{n-1}(X) \stackrel{d_n^{\varepsilon}}\leftarrow C_n(X) \leftarrow \cdots,
$$
where
$
d_n^{\varepsilon}(\sigma)=
\sum\limits_{i=1}^n (-1)^i \partial_i^{n, \varepsilon}(\sigma)$.

Homology groups $H_n^{\varepsilon}(X)$ of the complex called
{\em homology groups Goubault} semicubical set $X$.
They were studied in \cite{gou1995}.

Groups $H_n^0(X)$ are called {\em initial}, and $H_n^{1}(X)$ -- {\em final}
homology groups Goubault.

\begin{proposition}\label{cubdir}
Let $X= X_1\cup X_2$ -- union semicubical subsets $X_1\subseteq X$ and
$X_2\subseteq X$. Then for every $\varepsilon\in \{0, 1\}$ there is a long
exact sequence
\begin{multline*}
  0 \leftarrow H_0^{\varepsilon}(X) \leftarrow H_0^{\varepsilon}(X_1)\oplus
H_0^{\varepsilon}(X_2)
\leftarrow H_0^{\varepsilon}(X_1\cap X_2) \leftarrow \cdots \\
  \cdots \leftarrow H_n^{\varepsilon}(X) \leftarrow H_n^{\varepsilon}(X_1)\oplus
H_n^{\varepsilon}(X_2)
\leftarrow H_n^{\varepsilon}(X_1\cap X_2) \leftarrow \cdots \\
\end{multline*}
\end{proposition}
{\sc Proof} repeats almost verbatim the proof of Proposition \ref{exactcube}.

\subsection{Directed homology groups of the state category}

Homology groups of a small category $\mC$ with coefficients in the functor
 $F:\mC\rightarrow \Ab$ is defined as values $\coLim_n^{\mC}F$ of left
derived functors of $\coLim^{\mC}: \Ab^{\mC}\to \Ab$ on $F$.

Let $(S, M(E,I))$ be a state space, and $K(S)$ be its the state category.
Consider the functors $\Delta^0\ZZ: K(S)\to \Ab$ and $\Delta^1\ZZ: K(S)^{op}\to \Ab$,
taking on the objects the constant values $\Delta^0\ZZ(s)= \Delta^1\ZZ(s)= \ZZ$.
And on the morphisms defined by the formulas

$$
\Delta^{\varepsilon}\ZZ(s\stackrel{\mu}\to s')= \left\{
\begin{array}{cl}
1_s, & \mbox{ если } \mu=1,\\
0, & \mbox{ если } \mu\not=1,\\
\end{array}
\right.
$$

\begin{definition}
Directed homology groups of the state space defined by the formulas
%\begin{gather*}
$$
H_n^0(S,M(E,I))=_{def} \coLim_n^{K(S)}\Delta^0\ZZ, \quad
H_n^1(S,M(E,I))=_{def} \coLim_n^{K(S)^{op}}\Delta^1\ZZ.
%\end{gather*}
$$
For any elementary Petri nets $\mN$ its homology groups
$H_n^{\varepsilon}(\mN)$ defined as the homology groups of its state space.
\end{definition}

\begin{proposition}
For elementary Petri nets $\mN_n$ obtained from the pipeline Petri nets $\fP_n$ 
by deleting the transition $t_1$, homology groups equal
$$
H_k^{\varepsilon}(\mN_n)=\left\{
\begin{array}{cc}
\ZZ, & \mbox{ if } k=0\\
0, & \mbox{ if } k>0.
\end{array}
\right.
$$
\end{proposition}
{\sc Proof.} By Lemma \ref{subnet1}, the state category $\mC_n$ of
Petri net $\mN_n$ has an initial and terminal objects.
If a small category has a terminal object, then the colimit functor on the
category equals to the value of this functor on the terminal object.
Hence $H_k^{\varepsilon}(\mN_n)= \coLim_k\Delta^{\varepsilon}\ZZ=0$ for $k>0$ and $H_0^{\varepsilon}(\mN_n)= \coLim\Delta^{\varepsilon}\ZZ=\ZZ$.
\hfill $\Box$

According to \cite[Corollary 5]{X2012} for any 
linear order on $E$, there are isomorphisms

\begin{equation}
\label{spaceqube1}
\coLim_n^{K(S)}\Delta^0\ZZ \cong H_n^0(Q(S,E,I)), \quad
%\label{spaceqube2}
\coLim_n^{K(S)^{op}}\Delta^1\ZZ \cong H_n^1(Q(S,E,I)).
\end{equation}

Hence we get the following interpretation of directed homology for a state space in dimension
$0$. State $s$ is called to be a {\em deadlock}, if there is no $a\in E$
satisfying $s\cdot a\in S$. It is called a {\em sender}, if
there is no such a pair $s'\in S$ and $a\in E$ for which $s'\cdot a=s$.
In the state category, deadlock is an object $s$ which has not a morphisms $\alpha\not=1_s$ 
with $dom\alpha=s$. A sender $s$ has not $\alpha\not=1_s$ such that 
$cod\alpha=s$.

\begin{proposition}
The group $H_0^0(S, M(E,I))$ is isomorphic to the free abelian group generated
by deadlocks, and $H_0^1(S, M(E,I))$ generated by senders.
\end{proposition}
{\sc Proof.} It follows from \cite[Corollary 5]{X2012} 
that $H_0^0(S, M(E,I))$ is isomorphic to the cokernel of 
$d_1^0: LQ_1(S,E,I)\to LQ_0(S,E,I)$ defined as $d_1^0(s,a)=s$.
Cokernel is generated by the set obtained by the identification with $0$ of all $s\in S$ 
for which there are $a\in E$ satisfying $s\cdot a\in S$. It is removed from the set $S$
all objects which are no deadlocks. The remaining set generates the cokernel.
It is proved similarly that $H_0^1(S, M(E,I))$ generated by senders.\hfill $\Box$

\begin{example}\label{h0eps}
State category  of pipeline elementary Petri nets has no senders  or deadlocks,
hence $H_0^0(\fP_n)= H_0^1(\fP_n)=0$.
\end{example}

\subsection{Directed homology groups of pipeline}

According to the above example,  groups $H_0^0(\fP_n)$ and $H_0^1(\fP_n)$ equal
zero.
The following statement shows that $H_k^0(\fP_n)$ and $H_k^1(\fP_n)$ equal $0$ for all
$k\geq 0$.

\begin{theorem}
$H_k^{\varepsilon}(\fP_n)=0$, for all $n\geq 2$, $k\geq 0$ and $\varepsilon\in \{0,1\}$.
\end{theorem}
{\sc Proof.}
Above, we found that the semicubical set $Q(\fP_n)$
is equal to  union of semicubical sets $Q(\mN_n)$ and $Q(\mN'_n)$.
The state categories of $\mN_n$ and $\mN'_n$ have the greatest and the least elements.
This means that $H_k^{\varepsilon}(\mN_n)= H_k^{\varepsilon}(\mN'_n)=0$ for $k>0$, and
$H_0^{\varepsilon}(\mN_n)=H_k^{\varepsilon}(\mN'_n)= \ZZ$.
By the suggestion of \ref{union},  $Q(\mN_n)\cap Q(\mN'_n)\cong Q(\mN_{n-1})\sqcup Q(\mN_{n-1})$.
It follows that the exact sequence of the fragment suggests \ref{cubdir}
$$
\leftarrow H_{k-1}^{\varepsilon}(Q(\mN_n)\cap Q(\mN'_n))
\leftarrow H_k^{\varepsilon}(Q(\fP_n))\leftarrow H_k^{\varepsilon}(Q(\mN_n))\oplus H_k^{\varepsilon}(Q(\mN'_n)) \leftarrow
$$
leads to the equations $H_k^{\varepsilon}(Q(\fP_n))=0$ for $k\geq 2$.
In addition, there is an exact sequence
\begin{multline*}
0 \leftarrow H_0^{\varepsilon}(Q(\fP_n)) \leftarrow H_0^{\varepsilon}(Q(\mN_n))
\oplus H_0^{\varepsilon}(Q(\mN'_n)) \stackrel{H_0^{\varepsilon}(\theta)}\leftarrow \\
H_0^{\varepsilon}(Q(\mN_n)\cap Q(\mN'_n))\leftarrow H_1^{\varepsilon}(Q(\fP_n)) \leftarrow 0
\end{multline*}
By example \ref{h0eps} and isomorphisms (\ref{spaceqube1}), we have $H_0^{\varepsilon}(Q(\fP_n))=0$. 
We obtain an epimorphism
$\ZZ\oplus \ZZ \stackrel{H^{\varepsilon}_0(\theta)}\to \ZZ\oplus \ZZ$.
But $\ZZ\oplus \ZZ$ is projective object in the category of Abelian groups.
It follows that there exists a homomorphism $\gamma$ such
that $H_0^{\varepsilon}(\theta)\circ \gamma=1_{\ZZ\oplus \ZZ}$.
This means that the determinant of
$H^{\varepsilon}_0(\theta)$ is invertible where the kernel $H_0^{\varepsilon}(\theta)$ equally zero.
Consequently, $H_1^{\varepsilon}(Q(\fP_n)) = 0$. We have proved that
$H_k^{\varepsilon}(Q(\fP_n)) = 0$ for all $k\geq 0$.
 Isomorphisms (\ref{spaceqube1}) give
$H_k^{\varepsilon}(\fP_n)\cong H_k^{\varepsilon}(Q(\fP_n))$. 
\hfill $\Box$

\end{document}